\documentclass[%
 reprint,
 amsmath,amssymb,
 aps,
]{revtex4-2}

\usepackage{graphicx}% Include figure files
\usepackage{color}

\usepackage{dcolumn}% Align table columns on decimal point
\usepackage{bm}% bold math
\begin{document}

\preprint{APS/123-QED}

\title{Effect of substrate miscut angle on critical thickness, structural and electronic properties of MBE-grown NbN films on c-plane sapphire}% Force line breaks with \\

\author{Anand Ithepalli}
\author{Saumya Vashishtha}
\author{Naomi Pieczulewski}
\author{Qiao Liu}
\author{Amit Rohan Rajapurohita}
\author{Matthew Barone}
\author{Darrell Schlom}
\author{David A. Muller}
\author{Huili Grace Xing}
\author{Debdeep Jena}
\affiliation{Cornell University}

\date{\today}% It is always \today, today,
             %  but any date may be explicitly specified

\begin{abstract}

We report the structural and electronic properties of niobium nitride (NbN) thin films grown by molecular beam epitaxy on c-plane sapphire with miscut angles  of $0.5^\text{o}$, $2^\text{o}$, $4^\text{o}$, and  $10^\text{o}$ towards m-axis. X-ray diffraction (XRD) scans reveal that the full width at half maximum of the rocking curves around the 1 1 1 reflection of these NbN films decreases with increasing miscut. Starting from 76 arcsecs on $0.5^\text{o}$ miscut, the FWHM reduces to almost 20 arcsecs on $10^\text{o}$ miscut sapphire indicating improved structural quality. Scanning transmission electron microscopy (STEM) images indicate that NbN on c-sapphire has around 10 nm critical thickness, irrespective of the substrate miscut, above which it turns columnar.  The improved structural property is correlated with a marginal increment in superconducting transition temperature $T_\text{c}$ from 12.1 K for NbN on $0.5^\text{o}$ miscut sapphire to 12.5 K for NbN on $10^\text{o}$ miscut sapphire.

\end{abstract}

%\keywords{Suggested keywords}%Use showkeys class option if keyword
                              %display desired
\maketitle

%\section{Introduction}

Since the discovery of its superconductivity in 1941 \cite{NbNsuperconductivity}, niobium nitride has gained a lot of interest. Brauer identified the multiple structural phases, each with a characteristic Nb:N ratio, in the niobium-nitrogen binary phase diagram \cite{BrauerNbNxPhases}. The superconducting transition temperature of niobium nitride varies between 0.4 K $\leq T_\text{c} \leq $ 17.3 K, typically increasing with increasing N:Nb ratio \cite{NISTdatabase, JGWrightNbNx}. Wright \textit{et al}. discussed the growth windows of phase-pure niobium nitride films on c-plane sapphire substrate using plasma-assisted molecular beam epitaxy (PAMBE) \cite{JGWrightNbNx}. The rocksalt phase of niobium nitride, referred to as $\delta$-NbN, with a Nb:N stoichiometry around 1:1 has the highest $T_\text{c}$ among all the pure niobium nitride phases in bulk form \cite{NISTdatabase}. Wright \textit{et al}. confirmed that among phase-pure niobium nitride thin films, $\delta$-NbN has the highest $T_\text{c}$ around 13 K when grown on c-sapphire using PAMBE. Despite having the highest $T_\text{c}$, $\delta$-NbN showed 3-dimensional island growth mode and twinning due to the symmetry mismatch with the c-plane sapphire substrate, resulting in broad rocking curves (RCs) around the symmetric x-ray diffraction (XRD) peaks with full width at half maximum (FWHM) reaching 200 arcsecs \cite{JGWrightNbNx}. 

\begin{figure}
    \includegraphics[width=0.5\textwidth]{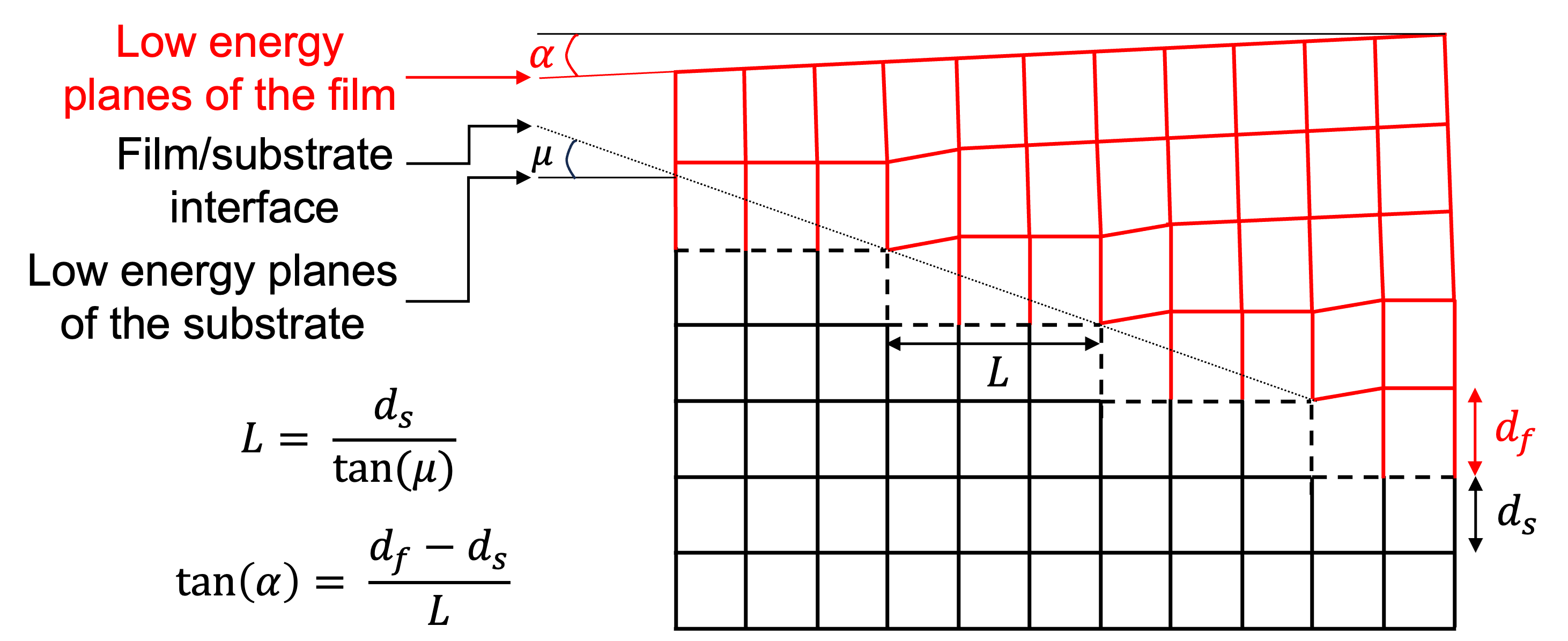}
    \caption{Schematic of the Nagai tilt. Substrates with a miscut angle $\mu$ form terraces with atomic steps on the surface to lower the surface energy. The terrace width $L$ depends on the miscut angle and the step height $d_s$ which is the inter-planar spacing of the low energy planes. The difference in the inter-planar spacings of the film $d_f$ and the substrate $d_s$ causes an epitaxial tilt $\alpha$ which is the angle between the low energy planes of the film and the substrate.}
    \label{fig:Nagai}
\end{figure}

In this report we attempt to answer the following question - can the structural quality of MBE-grown $\delta$-NbN on c-plane sapphire substrate be further improved? The use of high miscut substrates is identified as a possible solution. For instance, Nagai had shown improvement in the structural quality of In$_\text{x}$Ga$_\text{1-x}$As films grown on optimal miscut angles of GaAs substrates \cite{Nagai1974}. He discovered that the planes of the In$_\text{x}$Ga$_\text{1-x}$As film were tilted with respect to the planes of GaAs due to the difference in their inter-planar spacings as shown in Fig. \ref{fig:Nagai}. He attributed the structural improvement to the epitaxial tilt which reduces the nucleation degrees of freedom, due to the terrace widths being smaller than the nucleating grain sizes, resulting in a reduction in the twin and anti-phase domains. 

The epitaxial tilt is observed in many material systems since Nagai's report. Among nitrides, a recent example of Nagai tilt is shown in aluminum gallium nitride ternary alloy thin films grown on large miscut aluminum nitride substrates \cite{AlGaNonAlNmiscut}. Further, reduction of defects such as the surface hillocks on N-polar gallium nitride (GaN) grown on c-sapphire \cite{GaNonSapphireMiscut} and on N-polar GaN single crystals \cite{GaNonGaNMiscut} has been achieved by employing large miscut substrates. 

%\section{Experimental Methods}

\begin{figure}
    \includegraphics[width=0.45\textwidth]{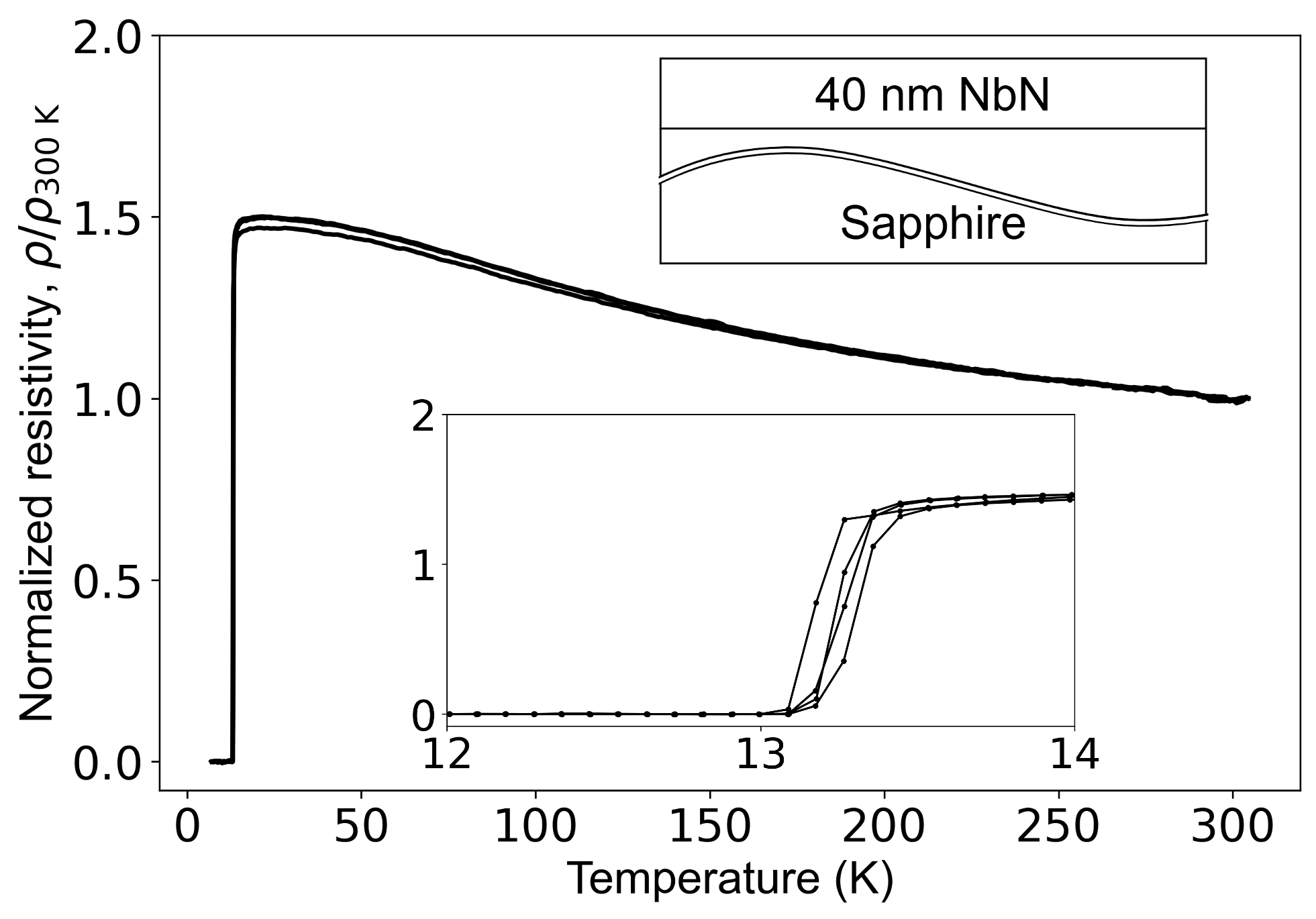}
    \caption{Normalized resistance vs temperature curve for all 4 NbN films on 0.5$^\text{o}$ miscut c-sapphire substrates. This control set has a $T_\text{c}$ spread of 0.2 K as opposed to 0.4 K spread in the set of NbN films grown on sapphire substrates with different miscuts.}
    \label{fig:RvT_0p5}
\end{figure}

Motivated by the Nagai tilt induced reduction of defects, $\delta$-NbN was deposited on large miscut c-sapphire substrates. We refer to $\delta$-NbN as NbN in the rest of this report. NbN was grown on c-sapphire with four different miscut angles of $0.5^\text{o}$, $2^\text{o}$, $4^\text{o}$, $10^\text{o}$ along the m-axis. Such large miscut substrates are difficult to obtain and for this study specially prepared wafers by Kyocera Inc. were used. To avoid experimental discrepancies due to growth-to-growth variations, all substrates with various miscut angles were co-loaded on a lapped silicon carrier wafer using indium mounting. Using a nitrogen plasma source and Nb from an E-beam source in a PAMBE system, around 40 nm NbN was then deposited on the co-loaded substrates at a substrate temperature of 600$^\text{o}$C measured by a thermocouple. The growth conditions are similar to that reported by Wright et al. \cite{JGWrightNbNx}. A control set of NbN grown on 4 identical c-sapphire substrates with $0.5^\text{o}$ miscut are also characterized and the key results from this control set are presented in the supplementary Fig. \ref{fig:2th-w-0p5}, Fig. \ref{fig:w_aligned_toNbN_0p5}, and  Fig. \ref{fig:RvT_0p5}. 

%\section{Results}

%\subsection{Single crystal XRD symmetric scans}

\begin{figure}
    \includegraphics[width=0.45\textwidth]{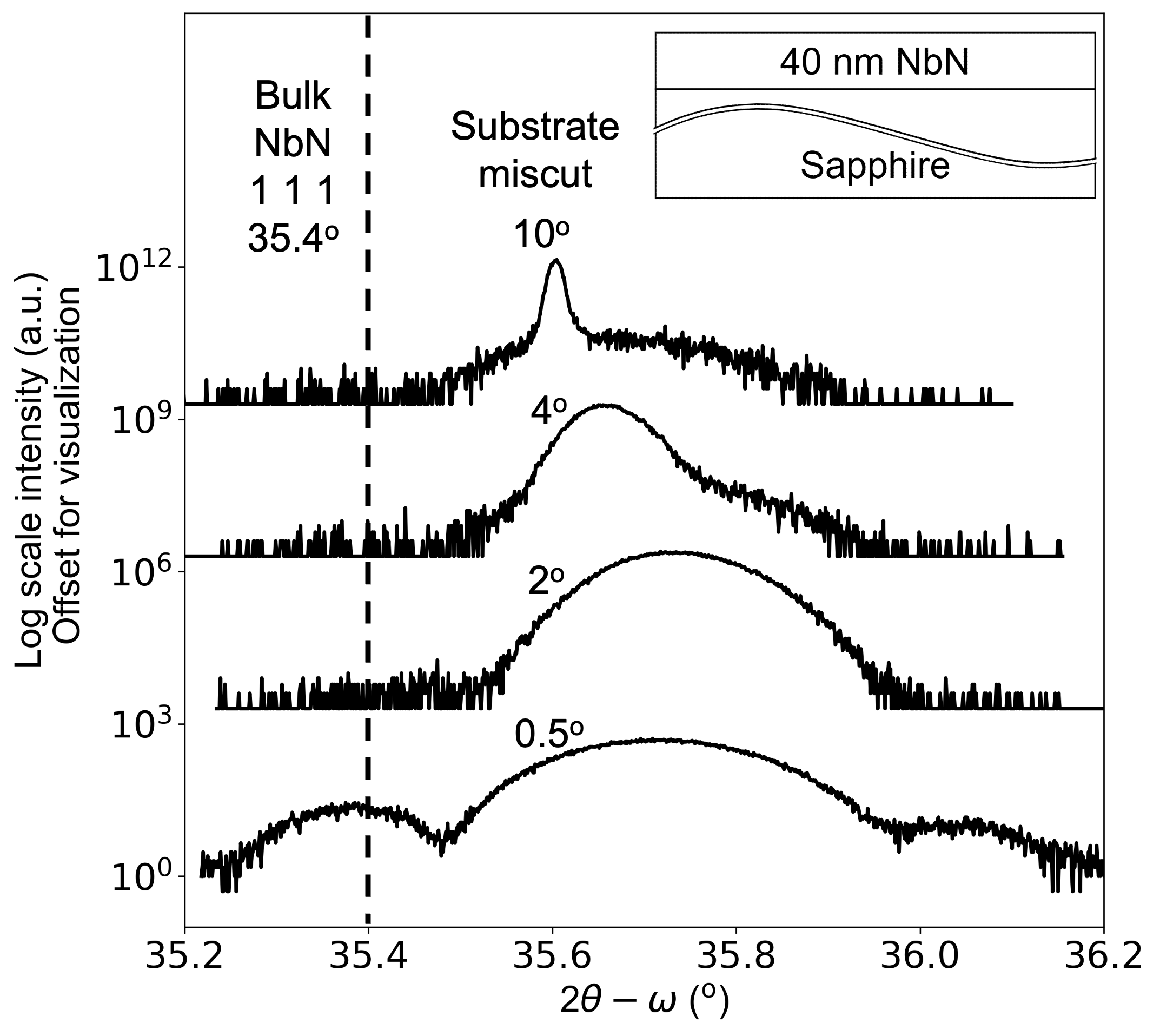}
    \caption{Symmetric $2\theta$-$\omega$ coupled scans around $2\theta=35.4^\text{o}$ expected for NbN 1 1 1 peak after aligning the instrument to NbN 1 1 1 reflection show narrower peaks with increasing miscut angle of the substrate and the peak position shifts closer towards the bulk NbN peak position.}
    \label{fig:2th-w}
\end{figure}

Coupled symmetric XRD scans around the 1 1 1 reflection of NbN on sapphire substrates with miscut angles of $0.5^\text{o}$, $2^\text{o}$, $4^\text{o}$ and $10^\text{o}$ are shown in Fig. \ref{fig:2th-w} along with the bulk NbN peak location at 2$\theta$ = 35.4$^\text{o}$. Interestingly, with increasing substrate miscut, the NbN 1 1 1 reflection peak width becomes narrower.  A shoulder peak with similar width persists across different miscuts, potentially due to the partially strained NbN below critical thickness. The FWHM of the primary $2\theta-\omega$ peaks around NbN 1 1 1 reflection reduces from 727.2\texttt{"} on $0.5^\text{o}$ miscut substrate to 57.6\texttt{"} on $10^\text{o}$ miscut substrate.

The broadening of $2\theta-\omega$ XRD peaks is typically attributed to crystallite/grain sizes and micro-strain following the Williamson-Hall equation \cite{WHequation}, $\beta\cos(\theta) = k\lambda/L+4\epsilon\sin(\theta)$, where $\beta$ is the FWHM of the diffraction peak at the diffraction angle $2\theta$, $k$ is the shape factor usually around 0.9, $\lambda$ is the wavelength of the x-ray source, $L$ is the crystallite size and $\epsilon$ is the strain. As is shown later in this manuscript using scanning transmission electron microscopy (STEM), the NbN columnar grain sizes are similar for substrate miscut of $0.5^\text{o}$ and $10^\text{o}$ whereas the $\beta$ changes by a factor of more than 12. Therefore, the Williamson-Hall equation reduces to $\beta \propto \epsilon$ implying that the strain in NbN grown on higher miscut substrate is lower. This is also evident by the fact that the primary NbN 1 1 1 reflection shifts closer to its bulk value as the substrate miscut is increased. One possible explanation for this reduction in strain is a reduction in nitrogen vacancies. Signatures of reduction in N-vacancies by employing a higher miscut substrate were also observed by Tatarczak et al. in gallium nitride homoepitaxial layers via photoluminescence measurements \cite{HenrykNvacancy}.

\begin{figure}
    \includegraphics[width=0.43\textwidth]{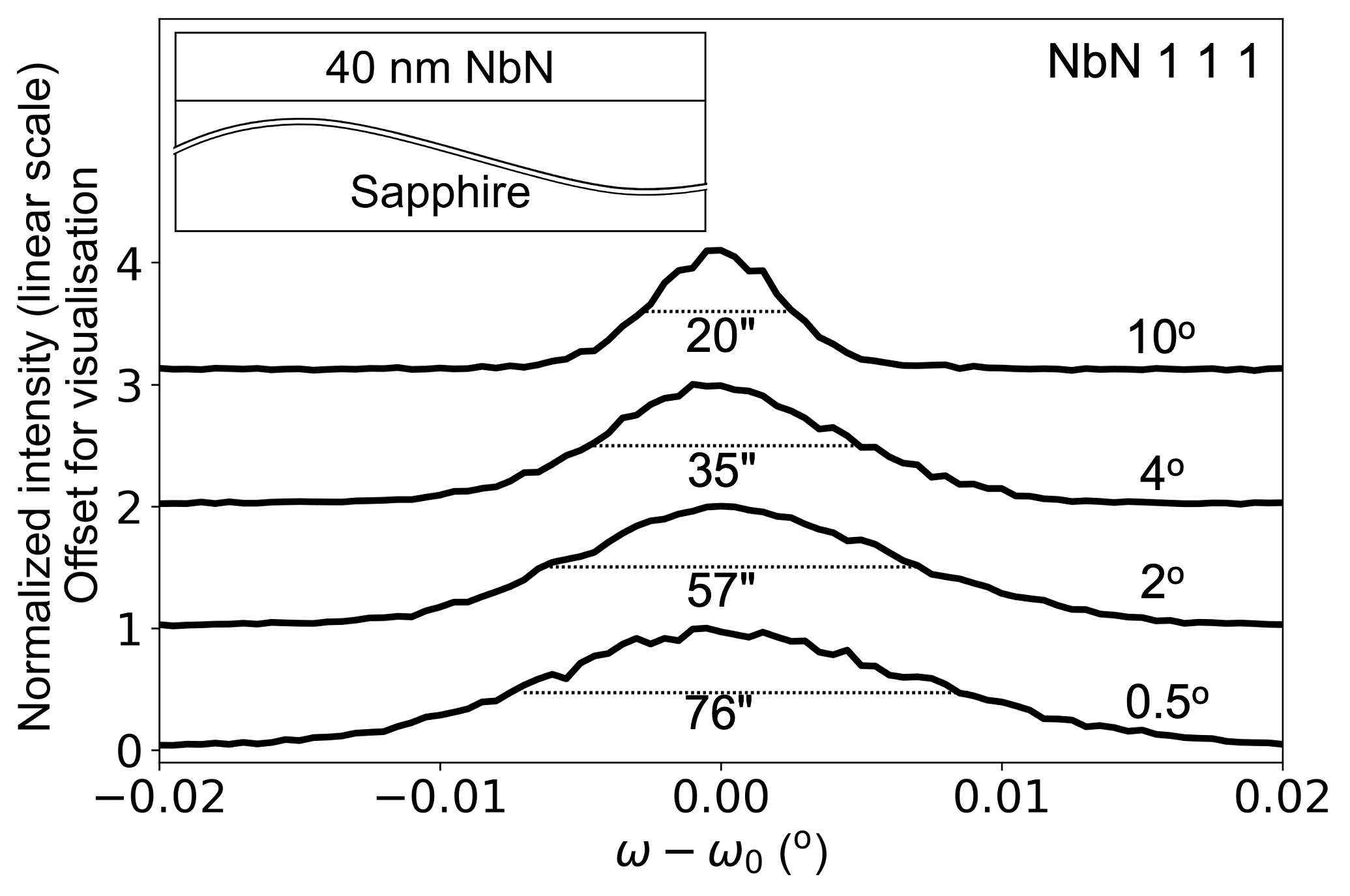}
    \caption{Normalized $\omega$ rocking curves (RCs) around the NbN 1 1 1 reflection show decreasing FWHM with increasing miscut angle of the substrate, the FWHM value of each RC is indicated in arcsecs (\texttt{"}) [3600 arcsecs (\texttt{"}) = 1 degree ($^\text{o}$)].}
    \label{fig:w_aligned_toNbN}
\end{figure}

\begin{figure*}
    \includegraphics[width=0.9\textwidth]{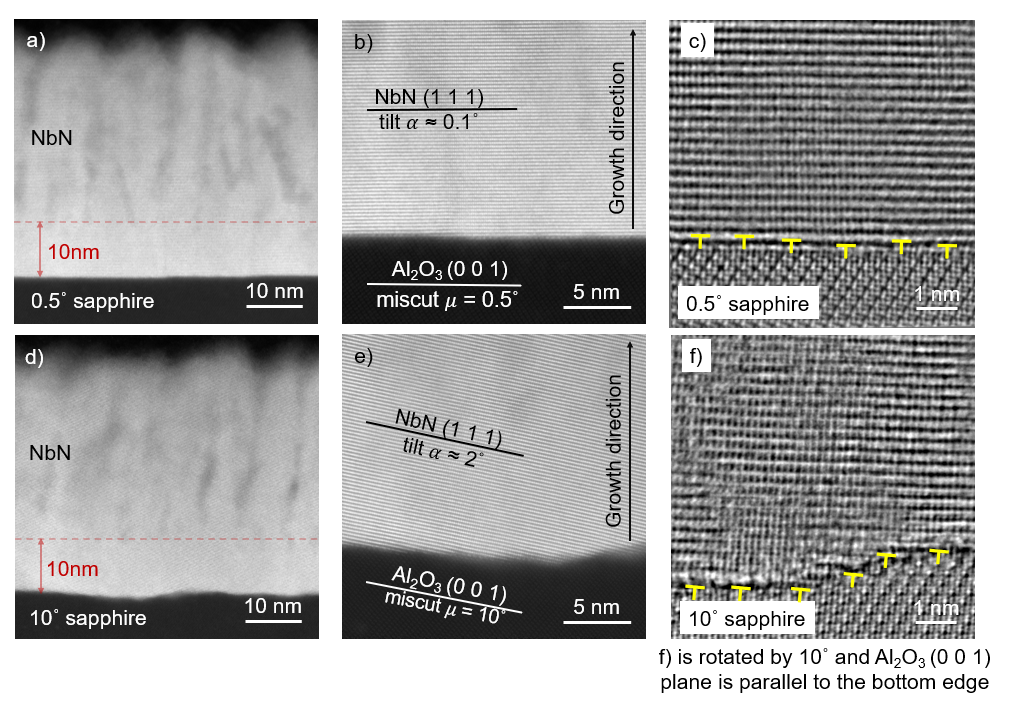}
    \caption{a), d) Wide field-of-view ADF-STEM images of NbN on 0.5$^\text{o}$ and 10$^\text{o}$ miscut sapphire, respectively. Both samples show a clean epitaxial NbN film in the initial 10 nm followed by columnar growth. b), e) Atomic resolution ADF-STEM images of the NbN to 0.5$^\text{o}$ and 10$^\text{o}$ miscut sapphire interface, respectively. The sapphire interface is aligned normal to the growth direction, and the NbN (1 1 1) planes prefer to be parallel to the sapphire (0 0 1) planes. c), f) Atomic resolution iDPC image of the NbN to 0.5$^\text{o}$ and 10$^\text{o}$ miscut sapphire interface aligning the sapphire (0 0 1) plane. The iDPC captures the tilts and disorder within the NbN that leads to blur in the image. The misfit dislocation between NbN and sapphire is noted in yellow across the interface. Both samples show a similar dislocation density $\sim$1.6$\times$10$^{12}$/cm$^2$ (assuming a 50 nm projection thickness). The high dislocation density reflects the large difference and lattice constant between the sapphire and relaxed NbN.}
    \label{fig:STEM}
\end{figure*}

\begin{figure*}
    \includegraphics[width=0.9\textwidth]{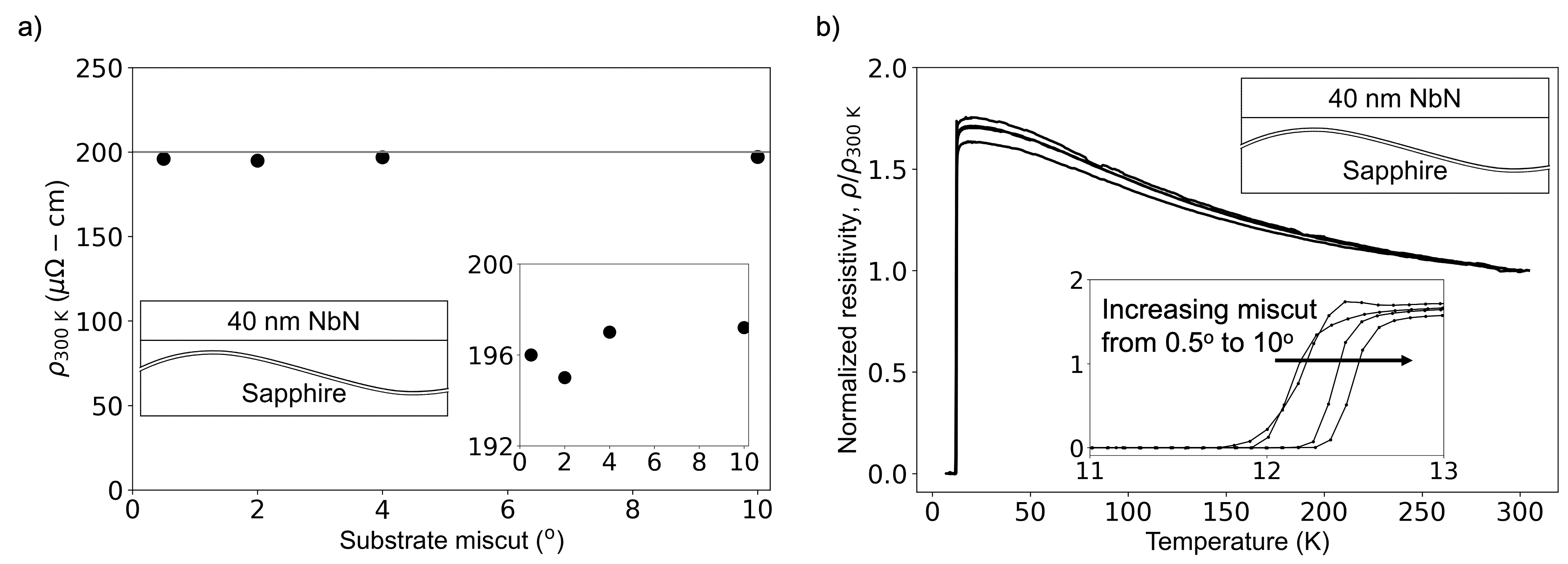}
    \caption{a) Room temperature resistivity of NbN films versus substrate miscut angle showing no impact of the substrate miscut. b) Marginal increment in NbN film's superconducting critical temperature, $T_\text{c}$, with increasing miscut angle of the sapphire substrate is seen from the normalized resistance vs temperature curve.}
    \label{fig:Tc_miscut}
\end{figure*}

%\subsection{Nagai tilt and apparent structural improvement}
Fig. \ref{fig:w_aligned_toNbN} shows the normalized and centered $\omega$ rocking curves (RCs) obtained after the XRD instrument is aligned to the (1 1 1) plane of NbN. FWHM values of symmetric RCs around NbN 1 1 1 reflection decrease monotonically from 76\texttt{"} on 0.5$^\text{o}$ miscut substrate to 20\texttt{"} on 10$^\text{o}$ miscut substrate indicating a reduction in defects such as threading dislocations and/or mosaicity. Quantifying these defects from STEM images is difficult due to the columnar grains.

%\subsection{Transmission electron microscopy}

Fig. \ref{fig:STEM}(a-c) show the STEM images of NbN on sapphire with 0.5$^\text{o}$ miscut at different magnifications, and Fig. \ref{fig:STEM}(d-f) for sapphire with 10$^\text{o}$ miscut. NbN was found to have around 10 nm critical thickness on c-sapphire irrespective of the substrate miscut as shown in Figs. \ref{fig:STEM}(a,d). Above this critical thickness, columnar growth of NbN is observed. The magnified images in Figs. \ref{fig:STEM}(b,e) prove the epitaxial and single crystal nature of the NbN film close to the NbN/sapphire interface within the critical thickness. The angular shift in the NbN (1 1 1) planes with respect to the NbN/sapphire interface in Figs. \ref{fig:STEM}(b,e) has two components - the substrate miscut $\mu$ and the Nagai tilt $\alpha$ as shown in Fig. \ref{fig:Nagai}. C-plane sapphire has an inter-planar spacing $d_s = 0.216$ nm and the NbN (1 1 1) planes have an inter-planar spacing $d_f \simeq 0.254$ nm. Therefore, according to the Nagai tilt model, for $\mu=0.5^\text{o}$, $\alpha=0.09^\text{o}$ and for $\mu=10^\text{o}$, $\alpha=1.78^\text{o}$. The values obtained from the STEM images, for $\mu=0.5^\text{o}$, $\alpha\simeq0.1^\text{o}$ and for $\mu=10^\text{o}$, $\alpha\simeq2^\text{o}$ match closely with the Nagai tilt model.

Misfit dislocations of density $\sim$1.6$\times$10$^{12}$/cm$^2$ (assuming a 50 nm projection thickness) are observed to form at the sapphire/NbN interface. This is consistent with the fact that the thickness of coherently strained NbN is $h_\text{c} \simeq$ 0.1 nm as estimated from the simplified Matthews-Blakeslee equation, $h_\text{c} = \frac{b}{4 \pi f (1+\nu)}\left[ \ln\left(\frac{h_\text{c}}{b}\right) + 1 \right]$ \cite{MATTHEWS1974118}. Burgers vector magnitude, $b=0.311$ nm for NbN, $f=11.7\%$ is the lattice mismatch between NbN and sapphire, and $\nu$ is the Poisson's ratio of NbN taken to be 0.33. The density and periodicity of these misfit dislocations is independent of the substrate miscut as denoted by the yellow dislocation symbols in Figs. \ref{fig:STEM}(c,f).

%\subsection{Electrical transport and temperature dependence}

The experimental verification of the actual mechanism leading to improved XRD metrics of NbN on large miscut sapphire requires further work. However, the effect of structural improvement on the transport properties is investigated in this study. Fig. \ref{fig:Tc_miscut}(a) is a plot of room temperature resistivity of the NbN films on miscut sapphire substrates measured in a Van der Pauw configuration. Despite the evidence of structural improvement with higher miscut substrate, all the films showed very similar room temperature resistivity with no apparent trend probably due to electron-phonon scattering limited transport at 300 K.

The temperature dependent resistivity shown in Fig. \ref{fig:Tc_miscut}(b) were normalized to the room temperature resistivity for each miscut. The inset in Fig. \ref{fig:Tc_miscut}(b) shows the superconducting transition temperature. The $T_\text{c}$ increases marginally by 0.4 K with increasing miscut. The temperature dependent resistivity curves of the control set shown in the supplementary Fig. \ref{fig:RvT_0p5} have less than 0.2 K variance in $T_{\rm c}$. This suggests that the $T_\text{c}$ increase with increasing substrate miscut is not an artifact. It must be noted that these measurements were done fresh after the growth and may not be reproducible after exposure to atmosphere over extended periods of time due to the possibility of impurity diffusion along column boundaries \cite{ImpurityDiffusion}.

From the inset of Fig. \ref{fig:Tc_miscut}(b), the residual resistivity ratio (RRR), $\rho_{300\text{ K}}/\rho_{13\text{ K}}$, increases marginally from 0.59 for NbN on 0.5$^\text{o}$ and 4$^\text{o}$ miscut to 0.62 for NbN on 10$^\text{o}$ miscut. An increase in RRR is a typical signature of reduced defects. An exception in the RRR trend is that of 2$^\text{o}$ miscut, with an RRR of 0.57, which cannot be explained based on the structural results presented in this study and may be due to competing scattering mechanisms.

%\section{Conclusion}

In conclusion, we employed large miscut (up to 10$^\text{o}$) sapphire substrates and observed an improvement in structural quality of NbN films as assessed by XRD. We have also observed that irrespective of the substrate miscut, around 10 nm of highly crystalline NbN grows on on c-plane sapphire. Beyond this thickness, the NbN films relax into columns. These columns have boundaries which can potentially allow impurities like O and C atoms to diffuse in from the surface. The cause for the structural improvement with the miscut angle is not fully understood and requires further work. Growth of transition metal nitrides which do not have columnar structure on c-plane sapphire can help understand which defects are reduced by employing large miscuts.

%\section{Discussion}

This study points to a number of avenues to pursue. The first is to compare the thick ($>$ 10 nm) and thin ($<$ 10 nm) NbN films epitaxially grown on c-sapphire in terms of transport and structural properties. The choice of miscut direction in this study was m-axis of sapphire. A different choice, say a-axis, of miscut direction might be worth studying to verify the universality of this study. Also, 32.4$^\text{o}$ miscut towards m-axis would refer to r-plane. The results from this study  also indicate that r-plane sapphire should be explored for epitaxial NbN growth.

%\section{Acknowledgments}

This work is supported by the AFOSR/LPS program Materials for Quantum Computation (MQC) as part of the EpiQ team under award number FA9550-23-1-0688 monitored by Dr. Ali Sayir of AFOSR and Dr. Erin Cleveland of LPS, and partially by an ONR Grant \# N00014-22-1-2633 monitored by Dr. Paul Maki. The authors acknowledge the use of facilities and instrumentation supported by NSF through the Cornell University Materials Research Science and Engineering Center DMR-1719875.

\bibliography{NbNmiscutSapphire}

\clearpage

\section{Supplementary}

\begin{figure}
    \includegraphics[width=0.45\textwidth]{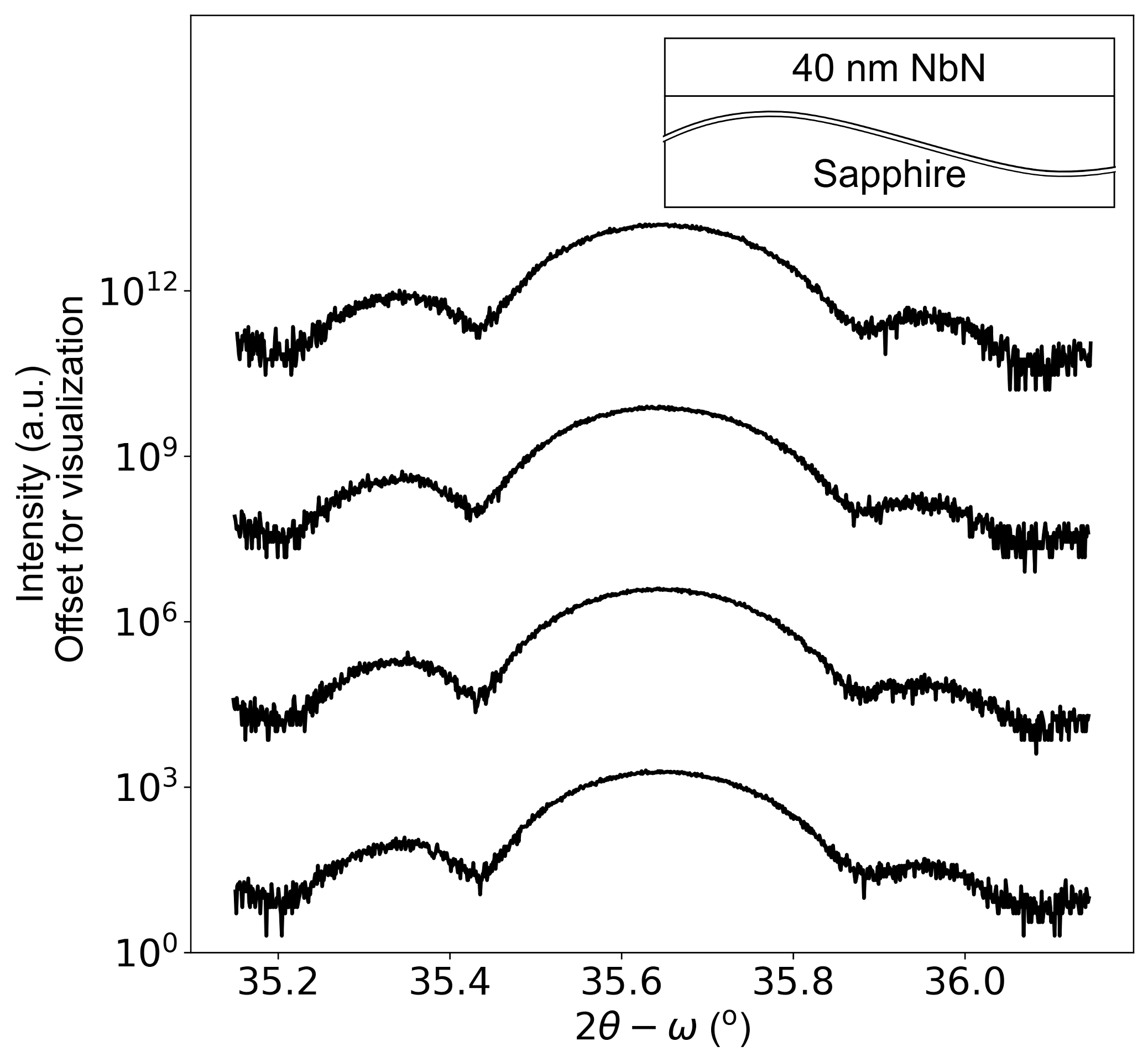}
    \caption{Symmetric $2\theta$-$\omega$ coupled scans around $2\theta=35.4^\text{o}$ expected for NbN 1 1 1 peak after aligning the instrument to NbN 1 1 1 reflection show very similar scans for all 4 NbN films on 0.5$^\text{o}$ miscut c-sapphire substrates.}
    \label{fig:2th-w-0p5}
\end{figure}

\begin{figure}
    \includegraphics[width=0.45\textwidth]{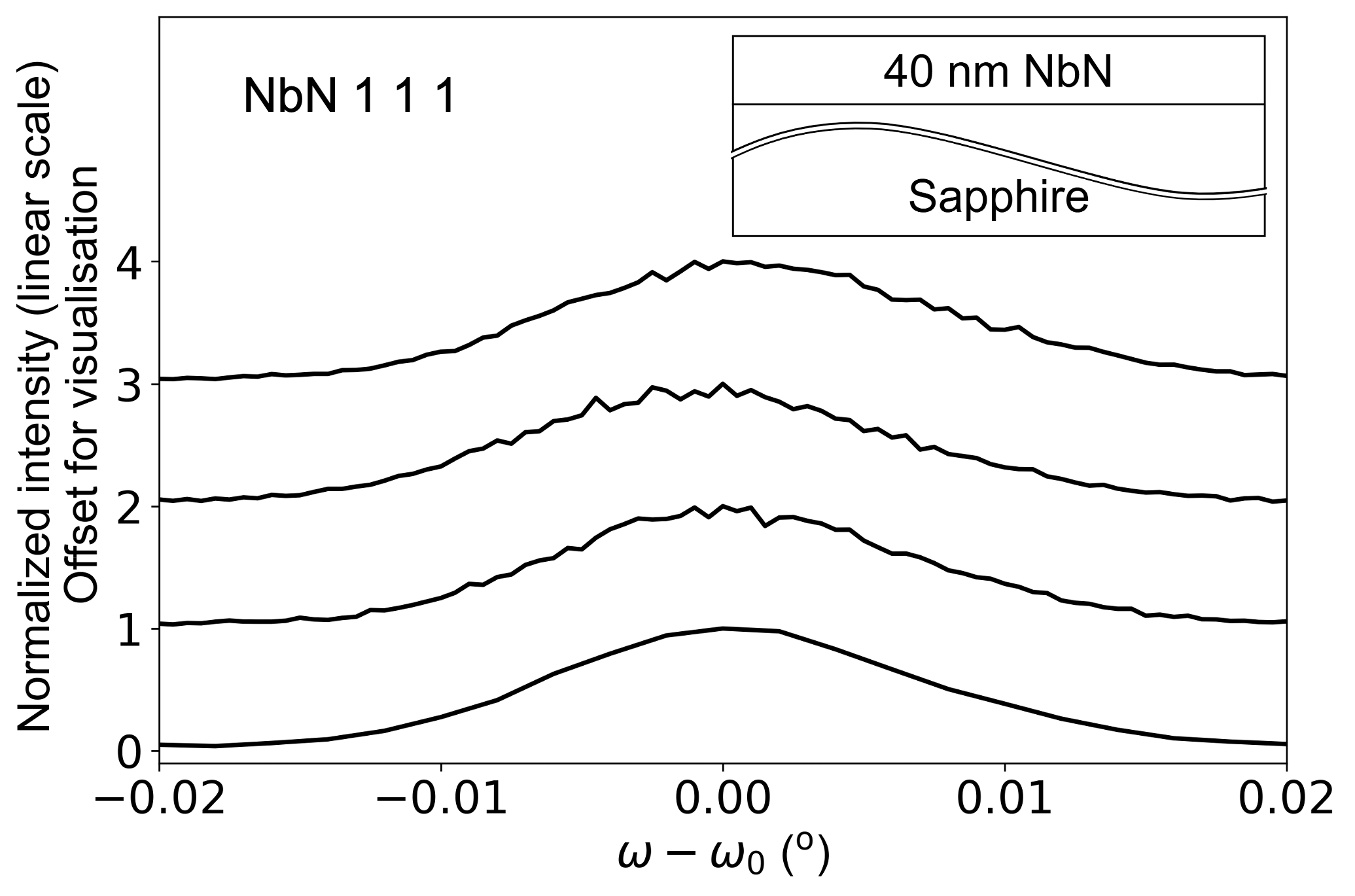}
    \caption{Normalized $\omega$ rocking curves (RCs) around the NbN 1 1 1 reflection show very similar FWHM for all 4 NbN films on 0.5$^\text{o}$ miscut c-sapphire substrates.}
    \label{fig:w_aligned_toNbN_0p5}
\end{figure}

\end{document}